\documentclass[aps,twocolumn,superscriptaddress]{revtex4-1}
\usepackage{graphicx}
\usepackage{amsmath}
\usepackage{dcolumn}
\usepackage{color}
\usepackage{epstopdf}
\usepackage[caption=false]{subfig}

\usepackage{float} 

\begin{document} 

\title{Plasmoid-mediated reconnection during nonlinear peeling-ballooning edge-localized modes}
\author{Fatima Ebrahimi}
\affiliation{Princeton Plasma Physics Laboratory, and Department of Astrophysical Sciences, Princeton University NJ, 08544}

\date{\today}
 
\begin{abstract}
Plasmoid-mediated reconnection is investigated for nonlinear Peeling-Ballooning Edge-Localized Modes (P-B ELMs). The formation of current sheets and the transition to 3-D current sheet instability is demonstrated through fully nonlinear resistive MHD simulations of P-B ELMs in DIII-D discharges. Large-scale axisymmetric current sheets, as well as small-scale poloidally extending current sheets, are formed as the coherent P-B ELM filaments nonlinearly evolve. It is observed that, at high Lundquist numbers, these current sheets break during a reconnection burst, i.e. a secondary exponential growth of intermediate modes followed by relaxation due to the suppression of P-B drive.                                              
\end{abstract}

\maketitle 

\textit{Introduction:} Free energy provided by steep gradients in the edge region of fusion plasmas and in natural plasmas, such as in the Earth's magnetotail, can trigger kinetic and MHD instabilities leading to eruptive events. Due to the nonlinear interaction of these instabilities, the nonlinear saturated state could be a turbulent state and in some cases could exhibit repetitive cycles. In tokamaks, edge
instabilities are triggered due to the local unfavorable magnetic curvature combined with steep gradients of density and temperature in the edge pedestal region. These instabilities including burst-like edge localized modes (ELMs), which are routinely observed in the high-performance operation regimes, could limit the global confinement and potentially contribute to plasma disruptions.   Coupled peeling
(current-driven) and ballooning (pressure-driven) ELM instabilities are known to be ideal-like MHD modes,  which have been demonstrated
both experimentally (see examples in DIII-D~\cite{evan_2006_nature,Diallo_2015}, NSTX~\cite{Maingi2009prl,Canik_nstx_rmp,ahn_2014_nstx}, C-Mod~\cite{Diallo_cmod}, MAST~\cite{kirk_mast} and KSTAR~\cite{kstar_2016_prl}) and theoretically~\cite{elm_connor,snyder2002,wilson2002,wilson2006,snyder2009,leonard2014}. Linear studies~\cite{ferraro2010,burke} as well as nonlinear reduced MHD modeling~\cite{huysmans2007,orain2015,cathey2020} have been used to model the peeling-ballooning characteristics for some H-mode experimental discharges.
Understanding the physics of edge instabilities, in particular as described by a self-consistent model of the nonlinear evolution, remains a challenging problem in magnetic fusion confinement. 
In this letter, using full extended MHD simulations, we investigate the nonlinear evolution of peeling-ballooning modes and the associated physics of fast magnetic reconnection. 

In
toroidal fusion plasmas, magnetic reconnection has also been demonstrated to be
critical in the nonlinear dynamics of many processes, such as sawtooth oscillations,
plasma disruption, plasma relaxation and magnetic self-organization in spheromaks~\cite{brown2006two} and RFPs~\cite{choiprl,ebrahimi2008}. It is believed that there are two types of trigger mechanism for forced and spontaneous magnetic reconnection, respectively. In forced magnetic reconnection, ~\cite{ji98,yamada2000} oppositely-directed field lines are brought together as the result of  electromagnetic forces. Alternatively, reconnection can be forced nonlinearly as a secondary effect due to an existing primary perturbation (so called driven reconnection). Magnetic reconnection could also be spontaneous due to slow-growing global current-driven resistive-tearing instabilities forming global islands~\cite{fkr,coppi76} or Alfvenic growing chains of plasmoids in a current sheet~\cite{biskamp86,tajima_shibata97,shibata2001,loureiro07,bhattacharjee09}, which  have been shown to cause a fast reconnection rate in the resistive MHD
model.~\cite{bhattacharjee09,huang2013,loureiro2012,ebrahimi2015plasmoids,comisso_grasso2016}  
Fast reconnecting plasmoids in a toroidal tokamak configuration have been shown during helicity injection in NSTX~\cite{ebrahimi2015plasmoids,ebrahimi16}, as well as in cylindrical plasmas during core sawtooth activity~\cite{gunter15}.
Plasmoid instability as a secondary effect due to flow-driven instabilities~\cite{Jarrett2021} has also been demonstrated in natural plasmas. In both tokamak core and edge regions, although slow magnetic reconnection via formation of global magnetic islands have been well studied, understanding of fast magnetic reconnection via secondary plasmoid formations remains limited.

In the scrape-off layer (SOL) region of a tokamak, the onset of edge reconnecting 3-D current-sheet plasmoid instabilities  ~\cite{ebrahimi2016dynamo,ebrahimi2017ELM} have been demonstrated to grow on the poloidal Alfvenic time scales and can break the current sheets.
It was found that coherent current-carrying filament (ribbon-like) structures
wrapped around the torus are nonlinearly formed due to nonaxisymmetric reconnecting current
sheet instabilities, the so called peeling-like ELMs. The cyclic nonlinear behavior of the low-n current-driven ELMs was explained via direct numerical calculations of the fluctuation-induced emf term of a current-carrying filament.~\cite{ebrahimi2017ELM}.  These earlier results exhibited a spontaneous (direct) onset of reconnecting instabilities for given SOL current sheets in the absence of pressure gradient. In collisional plasmas, in addition to free energy from current-gradient drive, 
resistive pressure-driven modes (with $\Delta{}' < 0$) in a cylinder~\cite{ebrahimi2002, paccagnella2013} as well as ballooning modes in the context of near-Earth magnetotail~\cite{zhu2017} have also been shown to exhibit 3D magnetic-reconnection characteristics. Here, to present a more complete picture of ELMs, by including the effect of pressure gradient, we investigate the nonlinear onset of 3D reconnecting instabilities in the tokamak SOL region.  
Through a detailed full nonlinear extended MHD study (with high toroidal and poloidal resolution), we uncover some of the fundamental physics of the trigger mechanism of ELMs. We identify three phases: 1) a linear phase when linearly unstable intermediate-n ballooning modes grow as well as the nonlinearly driven low-n peeling modes, 2) nonlinear saturation that is accompanied by a secondary nonlinear surge of multiple P-B modes and the nonlinear generation of zonal fields and currents, 3) the relaxation that occurs as the P-B modes nonlinearly suppress the edge current, as well as peeling it away.

\textit{Model:} The nonlinear evolution of peeling-ballooning modes is investigated using full extended MHD simulations of DIII-D discharges. We employ the NIMROD code~\cite{sovinec04}, which solves the 3-D, nonlinear, time-dependent, 
compressible, extended MHD equations
with a mesh of finite elements for the poloidal (R-Z)
plane and finite Fourier series for the toroidal direction.
\begin{eqnarray}
\label{eq:nimrod}
\frac {\partial \textbf B }{ \partial t } = 
- \nabla \times \textbf E + \kappa_{divb} \nabla \nabla . \textbf B\\
\textbf E = 
- \textbf V\times \textbf B + \eta \mu_0 \textbf J+ \frac{1}{ne}
\textbf J\times \textbf B + \frac{m_e}{ne^2}\frac{\partial \textbf J}{ \partial t}\\
\mu_0 \textbf J = \nabla \times \textbf B\\
\frac {\partial n }{ \partial t } 
 + \nabla. (n \textbf V) = \nabla . D \nabla n\\
\rho (\frac {\partial \textbf V }{ \partial t } 
 + \textbf V . \nabla \textbf V) = \textbf J \times
\textbf B - \nabla p - \nabla . \Pi \\
\frac{n}{(\Gamma-1)} (\frac {\partial T_{\alpha}}{ \partial t } 
 + \textbf V_{\alpha} . \nabla T_{\alpha}) = - p_{\alpha} \nabla .\textbf V_{\alpha} - \nabla . \textbf q_{\alpha} +Q_{\alpha} 
\end{eqnarray}
where  $p_{\alpha}= n_{\alpha} k T_{\alpha}$, $\textbf q_{\alpha}= -n_{\alpha} [(\kappa_{||} - \kappa_{\perp}) \hat b \hat b + \kappa_{\perp} \textbf I] \cdot \nabla T_{\alpha}$, $\alpha =e,i$; $\eta$  is the magnetic diffusivity,
 and $\kappa_{divb}$ and D are numerical magnetic-divergence and density diffusivities.

We employ equilibrium reconstructions using data from ELMing DIII-D discharges. Our first plasma equilibrium is based on a DIII-D discharge studied in Ref.~\cite{brennan2006} (case A).
Our second equilibrium is a DIII-D discharge 145701 at time 33~ms provided by Zhong Li~\cite{liao2016} (case B). Equilibrium reconstructed profiles and numerical grid are shown in Fig.~1. The spatial resolution is 40x60 fifth or sixth  order polynomials in the
poloidal plane, and 43 and 86 toroidal modes are used. We evolve the full MHD equations with an anisotropic pressure model and continuity equation, and use perpendicular and parallel thermal diffusivities of 25 and $2.5 \times 10^5$ $\mathrm{m^2/s}$, respectively. We use either temperature-dependent magnetic diffusivity
($\eta [m^2 /s] \sim T^{-3/2} [eV]$, Spitzer resistivity) based on the evolving toroidally averaged temperature, or constant magnetic diffusivity. Viscous dissipation is  introduced only through
the unmagnetized part of the stress tensor ($\Pi$), which is treated as 
$-\rho \nu \nabla^2 \textbf V$ or $-\rho \nu W$, where $\nu$ is the 
kinematic viscosity and W is the rate of strain tensor.
\\
\begin{figure}
  \includegraphics[]{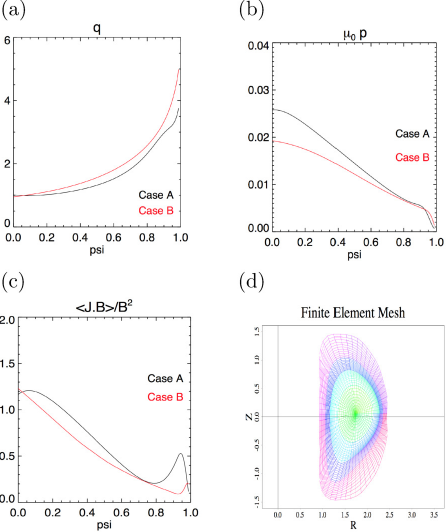}
\caption{Equilibrium (a) safety factor q, (b) pressure profiles, and parallel current density from two DIII-D discharges, case A~\cite{brennan2006} and case B (discharge 145701,~\cite{liao2016}) used in our nonlinear simulations. (d) finite element mesh used for case B.} 
  \label{fig:fig1} 
\end{figure}

\begin{figure}
  \includegraphics[]{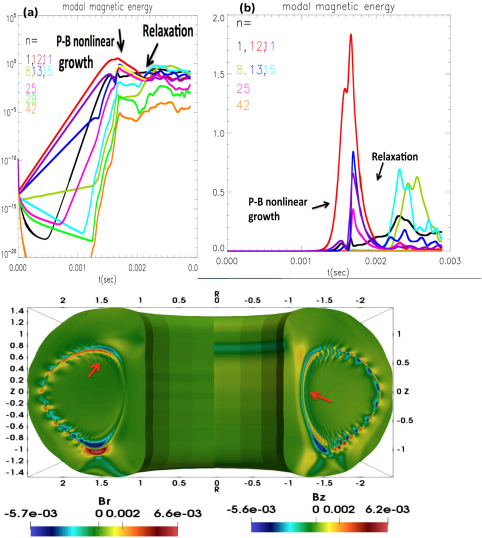}
  \caption{Modal magnetic energies vs. time during NIMROD simulations, (a) logarithmic (b) linear scales. The secondary sharp nonlinear growth of P-B modes followed by relaxation are shown. (c) Poloidal (R-Z) cross section of total perturbed radial (left) and vertical (right)  magnetic fields during nonlinear saturation at t = 1.68 ms. The arrows show the generation of poloidally extended zonal fields.  }
  \label{fig:fig2} 
\end{figure}

We first perform full nonlinear simulations for case A by evolving the set of equations (1-6) and starting with the
equilibrium reconstructed profiles shown in Fig.~\ref{fig:fig1} with temperature-dependent resistivity (equivalent to realistic  Lunquist number $S=\tau_R/\tau_A$).  The evolution of modal magnetic energies for various toroidal modes are shown in Fig.~\ref{fig:fig2}(a). Three sets of modes, low-n, intermediate and high-n are recognized. As it is seen, ballooning modes (with ballooning mode parity concentrated in the bad-curvature low-field side) with intermediate n  (n=10-14) first grow linearly,  while low-n current-driven peeling modes n=1-6 (with peeling parity),  which are linearly stable, later grow nonlinearly. High-n modes (n=20 and higher) grow nonlinearly due to mode-mode coupling.
Here, in our full nonlinear simulations,  two ELM cycles are completed. As the ballooning modes (with intermediate n) grow linearly and saturate, we find a secondary nonlinear growth of P-B modes (Fig.~\ref{fig:fig2}(b)). During this nonlinear growth phase,  axisymmetric zonal poloidal fields are further formed and amplified (at around t = 1.68 ms). These poloidal (R,z) magnetic fields are in fact nonlinear P-B mode structures that then  become locally more coherent and zonal (as shown with red arrows in Fig.~\ref{fig:fig2}). The formation of these zonal fields is due to the emf from the P-B modes themselves, which causes the relaxation of plasma current~\cite{ebrahimi2017ELM} and pressure at around t = 2~ms.

\begin{figure}
  \includegraphics[]{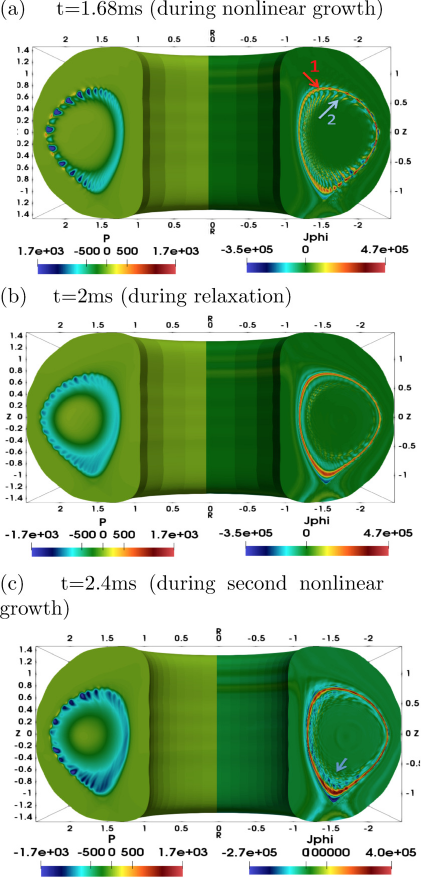}
\caption{Total perturbed (for all toroidal modes n=0-42) nonlinear pressure (left) and  toroidal current density (right) (a) after the secondary fast exponential growth of P-B modes at t = 1.65 ms (b) during the relaxation at t = 2 ms, and (c) during the nonlinear growth of the second cycle at t=2.4ms.}
  \label{fig:fig3} 
\end{figure}

\textit{Current sheet formation:} Figure~\ref{fig:fig3} shows total (including the contribution from all toroidal modes) current density during the nonlinear secondary growth. Two types of current sheets are identified during the nonlinear evolution. The first one is the poloidally extended axisymmetric current sheet (shown by arrow 1). According to Amp\'ere's Law, this toroidal axisymmetric current sheet is mainly driven by the poloidal zonal field generation ($B_r$ and $B_z$ shown in Fig.~\ref{fig:fig2}), given by
$ \mu_0 J_\phi = \frac{\partial B_r}{\partial z}-\frac{\partial B_z}{\partial r}$. The vertical variation of the generated radial zonal magnetic field,  $\frac{\partial B_r}{\partial z}$ (shown by red arrow on the left side of Fig.~\ref{fig:fig2}), combined with the radial variation of vertical magnetic field $\frac{\partial B_z}{\partial r}$  shown by red arrow on the right side of Fig.~\ref{fig:fig2}), produces an axisymmetric (n=0) toroidal current density shown in Fig.~\ref{fig:fig3} (a) (right side).
We should note that on the low-field side, this current sheet is mainly broken by the P-B modes during the secondary growth. The sign of the generated current sheet has an opposite sign of the equilibrium current edge, so it causes the suppression of the free energy (edge current) for the P-B modes.
 The second type of current sheet is the finger-type non-axisymmetic current sheets radially and vertically extending inside and out of the SOL (shown by the arrow numbered 2 in Fig.~\ref{fig:fig3}(a)).
 These latter current sheets further break and leave small-scale current blobs (plasmoids) as the modes nonlinearly saturate.

The total pressure from all the toroidal modes is also shown in Fig.~\ref{fig:fig3}~(a). Nonlinear small and larger scale finger-like structures in the outer edge region (low-field side with bad curvature) are also seen in Fig.~\ref{fig:fig3}~(a) (to the left) during saturation. It is thus observed that pressure fingers carry current. As the generated current sheets as well as pressure perturbations relax, the magnetic energy from P-B modes starts to decay as shown in Fig.~\ref{fig:fig1}~(b) at around t = 2~ms. As shown in Fig.~\ref{fig:fig3}~(b), the nonlinear structure of pressure and current density during the relaxation period shows the
 broadening of the edge pressure profile, as well as the decaying of the plasmoids (small-scale current sheets from P-B modes). The total current density relaxes to
 a mostly axisymmetric current density as is seen on the right in Fig.~\ref{fig:fig3}~(b). A second cycle of nonlinear P-B mode perturbations starts with finger-like pressure structures and small-scale current sheets as shown in Fig.~\ref{fig:fig3}~(c).

\begin{figure}
\includegraphics[]{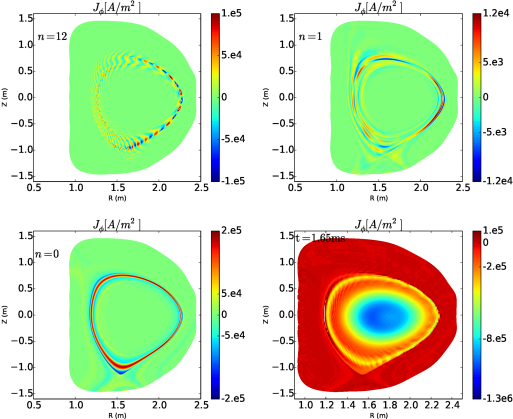}
\caption{Nonlinear mode structures of current density for (a) n=12 ballooning mode with localized structure on the low-field side (in the bad curvature region), (b) n=1 peeling mode with more poloidally symmetric structure, (c) nonlinearly generated axisymmetric n=0 and (d)  total current density  including the equilibrium (t=0) at t = 1.65 ms.} 
  \label{fig:fig4} 
\end{figure}

\textit{Modal decomposition and flows}: To better understand the nonlinear structure of the current density, we also investigate the modal decomposition of the total current shown in Fig.~\ref{fig:fig3}~(a). We separate the fields into mean and fluctuating components ($\textbf{J} =<\textbf{J}>_{n=0} +\tilde{\textbf{J}}_{n\neq 0}$), where the mean $<>$ is the toroidally ($\phi$) averaged axisymmetric field.  Figure~\ref{fig:fig4} shows the individual mode contributions of current density structures for several modes. The nonlinear ballooning mode structure with an intermediate n is shown in Fig.~\ref{fig:fig4}~(a), which shows current sheet fingers extending poloidally much further away from its simple linear structure (concentrated mostly in the bad curvature region). As can be seen low-n (n=1) mode structures are formed around the whole device (not limited to the low-field side).  The  current structures on the high-field side are mostly generated by the low-n modes (n=1 and 2). This is consistent with the peeling-mode structures as they extend near the edge on both sides of the device. As seen from the magnetic energy (Fig.~\ref{fig:fig2}a),  low-n modes (with peeling structures) are nonlinearly driven here.
Long-wavelength filaments on the high field side do have strong peeling components. The nonlinearly generated distinct axisymmetric n=0 current density  is also shown (Fig.~\ref{fig:fig4}c). This axisymmetric current density is nonlinearly generated to suppress and relax the edge current (i.e. with a different sign than the background current density).  Finally, all  the toroidal modes (n=0 and above) do nonlinearly contribute to affect the equilibrium current density
and suppress the edge current as well as  peeling away the edge current as seen in Fig.~\ref{fig:fig4}d.

\begin{figure}
\includegraphics[width=3.2in,height=2.in]{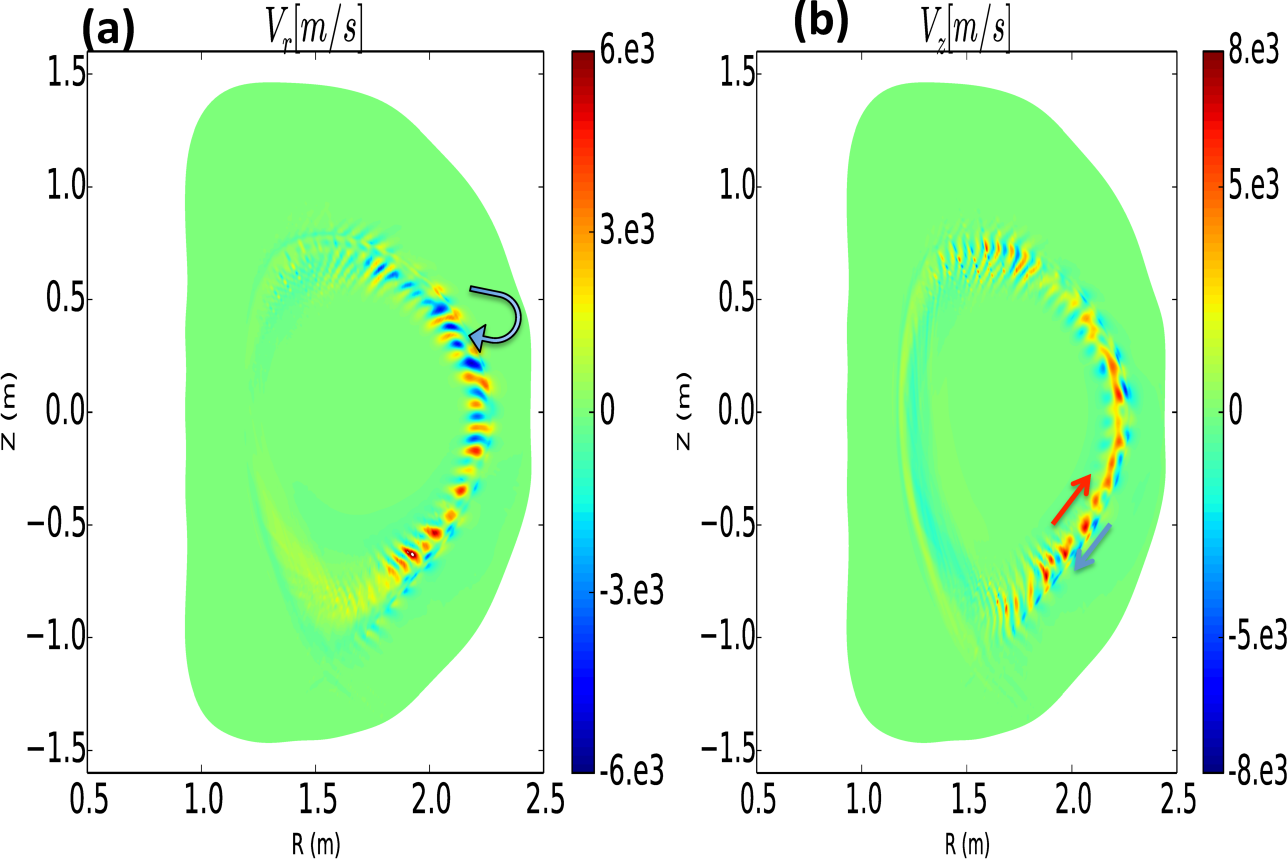}
\caption{Nonlinear poloidal flows, (a) radial (b) axial, generated  during
nonlinear onset of P-B modes during 2-fluid simulations. Arrows show vortices formation.} 
  \label{fig:fig5} 
\end{figure}

In addition to current sheets, edge flow vortices are also formed nonlinearly in the MHD regime. The generation of flows has been further verified in the the 2-fluid regime in simulations  by including the generalized Ohm's law (Eq.~2) but with a fixed magnetic diffusivity (equivalent to $S \sim 5 \times 10^4$) with 86 modes.
Poloidal flow vortices do form and are amplified  during the nonlinear growth of P-B modes as shown in Fig.~\ref{fig:fig5}. The source of these  vortices could be 
local reconnection along the current sheets as they break. As the local Alfven velocity, based on the poloidal zonal ($V_{A(pol)} = B_{(pol)}/\sqrt{\mu_0 \rho}$ with $B_p\sim 0.0065$ T, and $n=5\times 10^{19} m^{-3}$) is about 14~km/s, these nonlinear poloidal flows are also in the range of the poloidal Alfven velocity. However, in addition to a direct local reconnection effect,  the nonlinear Maxwell and Reynolds stresses could also contribute to generation of these flows.~\cite{ebrahimi2007} 
However, further analysis is required for the detailed dynamics of these flows.

\begin{figure}
\includegraphics[]{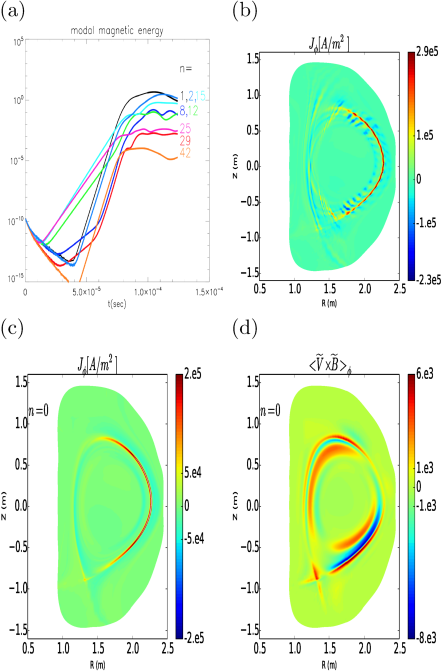}
\caption{Case B: (a) modal magnetic energies, (b) total toroidal current density (c) n=0 generated current density at t=0.11ms, and (d) poloidal cross section of EMF at t=0.11ms.} 
  \label{fig:fig6} 
\end{figure}
Lastly, to further investigate the robustness of the nonlinearly generated axisymmetic current sheets and the reconnection dynamics, we also  perform nonlinear simulations for another DIII-D discharge (Case B from Ref.~\cite{liao2016}) with uniform magnetic diffusivity (with S $\sim 10^5$).
Similarly, intermediate-n P-B modes grow linearly, while low-n peeling (current-driven) modes grow nonlinearly and saturate. 
In particular,  n=1 modes saturate  at large amplitude, and  higher-n P-B modes saturate at much lower amplitudes. Modal magnetic energy is shown in Fig.~\ref{fig:fig6}. The nonlinear dynamics of current density and zonal magnetic field generation is similar to Case A (Fig.~\ref{fig:fig2}). In this case as well,  two types of current sheets of poloidally extending (R,Z) (type 1) and radially extending blobs of current (type 2) are generated as the modes saturates (current density profile shown in Fig.~\ref{fig:fig6}b ). The axisymmetric current density profile, which is nonlinearly generated by the P-B modes to suppress source of instability itself, is similarly shown in Fig.~\ref{fig:fig6}. To further investigate the the relaxation of P-B modes, we calculate the toroidal fluctuation induced emf term in the toroidal direction $<\tilde{\textbf{V}} \times \tilde{\textbf{B}}>_{\phi} = 1/2Re[\widetilde V_r \widetilde B_z^{*} - \widetilde V_z \widetilde B_r^{*}]$. The toroidally averaged poloidal cross section of  the toroidal emf exhibits a  bi-directional structure, which is consistent with formation of axisymmetric
toroidal current density.\\

In summary, in the framework of the extended MHD model, we have presented a detailed fully nonlinear (and self-consistent) study of P-B ELMs. We find that as the linearly unstable intermediate-n ballooning modes and the nonlinearly driven peeling-type low-n modes  grow and saturate, it is during a secondary sudden nonlinear growth of P-B modes that nonlinear finger-like structures of ballooning modes are expelled into the outer edge region.  Radial (and poloidal) current expulsion occurs in the form of 3D plasmoids in the region of the SOL during a bursty reconnection process (a sudden growth followed by relaxation).  
Through modal decomposition of peeling and ballooning components of ELMs, we have also uncovered nonlinearly generated axisymmetic current sheets that suppress edge peeling drive and lead to relaxation of ELMs.
In our self-consistent ELM calculations, we show that the plasma edge goes through magnetic self-organization. As the P-B ELMs are suppressed, the plasma current and pressure are relaxed, but they grow again during a second cycle. Simulations with kinetic physics, such as further two-fluid modeling with ion gyro-viscosity effects and gyrokinetic modeling~\cite{dong2019} will be investigated in the future. We should recognize that as more experimental measurements demonstrate some direct signatures of magnetic reconnection including Whistler-frequency emissions~\cite{kim2020} and flow dynamics~\cite{lampert2021} during ELMs, nonlinear edge simulations could strongly contribute to the understanding of the driving mechanism of such dynamics and emissions. Our nonlinear extended MHD  P-B simulations of DIII-D, along with earlier simulations of SOL currents~\cite{ebrahimi2017ELM}  in NSTX, highlight the importance of magnetic reconnection and the associated plasma current relaxation in the edge region of tokamaks.  Filaments emerging from nonlinear peeling-ballooning instabilities display strong current sheets leading to the formation of plasmoids and bursty reconnection dynamics. 

\acknowledgements 
The author acknowledges Amitava Bhattacharjee for useful discussions and for reviewing the initial draft of this paper. We also acknowledge G. Dong, Z. Lin and D. Brennan for providing DIII-D eqdsk files. Computations were performed at NERSC. This work was supported by DOE grants DE-SC0010565, DE-AC02-09CHI1466, and DE-SC0012467.


\end{document}